\documentclass[preprintnumbers,twocolumn,prl,aps,superscriptaddress,footinbib,amsfonts,amsmath,amssymb,bm,showpacs,floatfix]{revtex4-1}

\makeatletter
\renewcommand*{\@fnsymbol}[1]{\ensuremath{\ifcase#1\or *\or \dagger\or \ddagger\or
   \mathsection\or \mathparagraph\or \|\or **\or \dagger\dagger 
   \or \ddagger\ddagger 
   \or \mathsection\mathsection
   \or \mathparagraph\mathparagraph
   \or \|\|
   \else\@ctrerr\fi}}
\makeatother

\usepackage[colorlinks=true,linkcolor=blue,citecolor=blue,urlcolor=blue]{hyperref}

\usepackage[T1]{fontenc}
\usepackage{blindtext}
\usepackage{amsmath}
\usepackage{amssymb}
\usepackage{graphics,graphicx}
\usepackage{bm}
\usepackage{epsfig}
\usepackage[normalem]{ulem}
\usepackage[usenames]{color}
\usepackage{mathbbol}
\usepackage{epstopdf}
\usepackage[utf8]{inputenc} 

\usepackage{makecell}

\allowdisplaybreaks

\newcommand{\nn}{\nonumber}

\renewcommand{\(}{\left(}
\renewcommand{\)}{\right)}

\begin{document}

\preprint{IPARCOS-UCM-25-050}
\preprint{DESY-25-128}  

\title{Determination of quark-gluon-quark interference within the proton}

\author{Alexey Vladimirov}
\thanks{\href{mailto:alexeyvl@ucm.es}{alexeyvl@ucm.es}}

\author{Guillermo Portela}
\affiliation{Departamento de Física Teórica and IPARCOS, Universidad Complutense de Madrid, \\ Plaza de Ciencias 1, 28040 Madrid, Spain}

\author{Simone Rodini}
\affiliation{Deutsches Elektronen-Synchrotron DESY, Notkestr. 85,22607 Hamburg, Germany}
\affiliation{Dipartimento di Fisica ``Alessandro Volta'', Universit\`a degli Studi di Pavia, 27100 Pavia, Italy}

\begin{abstract}

Quarks and gluon, as quantum particles, are subjects to various effects that go beyond the naive parton picture and are not captured by ordinary parton densities. In this work, we investigate the twist-three parton distribution functions, which encode quantum interference between quark–gluon–quark states, and for the first time, determine them directly from experimental data. The analysis combines observables described by collinear and transverse-momentum–dependent factorization theorems within a unified global fit, incorporating a complete leading-order QCD evolution at the twist-three level. The extracted distributions reveal a clear flavor-dependent patterns and distinct from zero at a statistically significant level ($2-3\sigma$). These findings provide the first quantitative evidence for quark–gluon–quark correlations within the proton, revealing its genuinely quantum nature and opening a new direction for precision studies of partonic correlations.

%Quarks and gluon as being quantum particles are subjects to various effects that go beyond the naive parton picture and are not present in ordinary parton densities. In this work, we consider the parton distribution functions of twist-three, that present quantum interference between quark-gluon-quark states, and for the first time, determine these distributions from the data, establishing their presence at a 2$\sigma$-3$\sigma$ significance. To achieve this result, we perform a global analyses of a novel kind, which combines distinct observables described by collinear and transverse momentum dependent factorization theorems melded by a complete implementation of QCD evolution at twist-three level. The extracted distributions reveal flavor-dependent patterns, demonstrating the genuinely quantum nature of the proton and opening a new direction for precision studies of partonic correlations.

%Quantum Chromodynamics predicts that protons are shaped not only by parton densities but also by quantum interference between quarks and gluons, which is encoded within the set of twist-three quark-gluon-quark distributions. By combining distinct observables with a complete implementation of QCD evolution, we achieve the first determination of said twist-three distributions, establishing their presence at a 2$\sigma$-3$\sigma$ significance. The extracted distributions reveal flavor-dependent patterns, demonstrating the genuinely quantum nature of the proton and opening a new direction for precision studies of partonic correlations.
\end{abstract}

\maketitle 

%%%%%%%%%%%%%%%%%%%%%%%%%%%%%%%%%%%%%%%%%%%%%%%%%%%%%%%%%%%%%%%%%%%%%%%%%%
{\it Introduction --- } Quarks and gluons are bound together into hadrons. Although the dynamics of their microscopic interactions is described by QCD \cite{Gross:2022hyw} and is well understood locally, the general picture of this binding remains a puzzle, and could not be rigorously described without the solution of the confinement phenomenon. Still, one can consider the QCD component of high-energy scattering avoiding direct discussion on the confinement problem, encapsulating instead nonperturbative parton dynamics into certain matrix elements, collectively known as parton distribution functions (PDFs), and determining them from the experimental data. This is a very productive and precise approach, built on the solid mathematical foundation of QCD factorization theorems \cite{Collins:1989gx}. By investigating PDFs, we have learned a lot about the proton's inner structure, and each extracted PDF brings us closer to a complete understanding of the hadronic internal composition.

In this work we present the proof-of-concept study that, for the first time, determines PDFs of twist-three. They represent matrix elements of three-point operators, and are interpreted as the interference between the two- and one-parton components of the hadron's wave function \cite{Politzer:1980me, Ellis:1982cd, Jaffe:1996zw}. Due to it, twist-three PDFs are intrinsically different from all those that are already known (see f.i. reviews~\cite{Gao:2017yyd, Angeles-Martinez:2015sea, Diehl:2003ny}), which generally fall into the class of twist-two PDFs, and are interpreted as parton densities. In this way, our work uncovers a new layer of parton dynamics, and explicitly demonstrates the truly quantum nature of non-perturbative QCD.

Despite there are many observables sensitive to twist-three effects, data-driven studies of twist-three distributions remain virtually absent. The problem is that a twist-three PDF is a function of two variables and each observable is sensitive only to a subregion or an integral over subregion of this function, making impossible a reconstruction of its complete shape. Furthermore, many observables have contributions of twist-two PDFs, which obfuscate the twist-three effects. As a consequence, there have been only a few attempts to determine separate elements of the twist-three PDFs, such as the renowned structure function $g_2$ \cite{Sato:2016tuz, Cocuzza:2025qvf}, or the, so-called, Efremov-Teryaev-Qiu-Sterman function \cite{Efremov:1984ip, Qiu:1991pp, Bury:2021sue}.

The key lies in the simultaneous analysis of different classes of observables and an accurate implementation of QCD evolution. In this way, each observable constrains a different aspect of the distribution, while the evolution equations tie together different regions and flavors, introducing important correlations between distant points. Together, these links form a skein of connections that capture the underlying behavior, even when the constraint on each individual element is weak.

%Following this strategy, we have performed the joint fit of observables of four different kinds: the structure function $g_2$, measured in polarized deep-inelastic scattering (DIS) \cite{Jaffe:1989xx, Kodaira:1994ge}; the moment $d_2$ determined in polarized DIS alongside $g_2$ and directly calculable in lattice QCD \cite{Burger:2021knd}; the single-spin asymmetry $A_{UT}^{\sin(\phi_h-\phi_s)}$ in semi-inclusive DIS (SIDIS), which probes twist-three effects through the Sivers function $f_{1T}^\perp$ \cite{Sivers:1989cc}; and double-spin asymmetry $A_{LT}^{\cos(\phi_h-\phi_s)}$ in SIDIS, where twist-three physics enters via the worm-gear-T function $g_{1T}^\perp$, also known as the Kotzinian-Mulders function \cite{Kotzinian:1995cz}. The analyses is done using the complete implementation of twist-three evolution at leading order (LO) \cite{Rodini:2024usc}. It must be noted that this is the first example of a joint consideration of processes of such diverse kinds, as well as the first practical test of the twist-three evolution. The success of this study further establishes the predictive power of QCD, and opens an avenue for global analyses of hadron structure at a larger scale.

Following this strategy, we have performed the joint fit of observables of four different kinds: the structure function $g_2$ in deep-inelastic scattering (DIS); the moment $d_2$; the single-spin asymmetry $A_{UT}^{\sin(\phi_h-\phi_s)}$ in semi-inclusive DIS (SIDIS); and double-spin asymmetry $A_{LT}^{\cos(\phi_h-\phi_s)}$ in SIDIS. The analyses is done using the complete implementation of twist-three evolution at leading order (LO) \cite{Rodini:2024usc}. It must be noted that this is the first example of a joint consideration of processes of such diverse kinds, as well as the first practical test of the twist-three evolution. The success of this study further establishes the predictive power of QCD, and opens an avenue for global analyses of hadron structure at a larger scale.

{\it PDFs of twist-three --- } The quark PDFs of twist-three are defined as following \cite{Braun:2009mi, Rodini:2024usc}
\begin{eqnarray}
&&\langle p,s|g \bar q(z_1n)F^{\mu+}(z_2n)\gamma^+q(z_3n)|p,s\rangle\\\nn &&= 2 \epsilon^{\mu\nu}_Ts_\nu(p^+)^2 M\int [dx] e^{-i p^+\sum_i z_i x_i}T(x_1,x_2,x_3),
\\
&&\langle p,s|g \bar q(z_1n)F^{\mu+}(z_2n)\gamma^+\gamma^5q(z_3n)|p,s\rangle\\\nn &&= -s^\mu_T(p^+)^2 M\int [dx] e^{-i p^+\sum_i z_i x_i}\Delta T(x_1,x_2,x_3),
\end{eqnarray}
where $p$ and $M$ are the momentum and mass of the hadron, $s^\mu$ is a spin vector, $g$ is the QCD coupling constant, $q$ is the quark field and $F^{\mu\nu}$ is the gluon field-strength tensor. The vector $n$ is a light-cone vector \footnote{We use the standard convention in the notation for the components of a vector in the light-cone decomposition. For a generic vector $v^\mu$ one defines
$$v^\mu=v^+\bar n^\mu+v^- n^\mu+v_T^\mu,$$
where $n^2=\bar n^2=0$, and $(n\bar n)=1$. The letter $T$ denotes the transverse components. The transverse Levi-Civita tensor is defined as $\epsilon_T^{\mu\nu}=\epsilon^{\mu\nu-+}$ with $\epsilon^{0123}=+1$.
}. The straight gauge links connect all fields successively, securing the gauge invariance, are not presented. 

\begin{figure}[t]
\begin{center}
\includegraphics[width=0.38\textwidth]{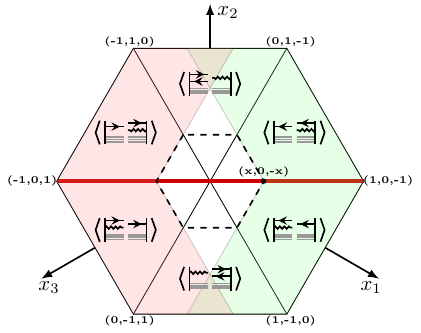}
\caption{\label{fig:hex} The domain of twist-three PDFs. Each sector corresponds to an interference between particular partonic configurations, shown in diagrams (with wavy lines being gluons, and straight lines being quarks). Different physical observables are sensitive to various sub-regions of the hexagon shown in colors and explained in the text.}
\end{center}
\end{figure}

The variables $x_{1,2,3}$ represent the momentum fractions carried by the fields and are restricted by causality as $-1<x_i<1$ \cite{Jaffe:1983hp}. The momentum conservation requires $x_1+x_2+x_3=0$, and thus reduces the number of independent variables to two. These constraints define the integration measure $[dx]=\delta(x_1+x_2+x_3)dx_1dx_2dx_3$, and the domain of the definition of twist-three distributions, which is conventionally presented in barycentric coordinates and has the shape of a hexagon. This representation is particularly convenient for manipulations with twist-three PDFs, because each sector of the hexagon represents a specific interference term defined by the sign pattern of $(x_1,x_2,x_3)$. In this context, positive values of $x_i$ correspond to a parton production process, while negative values indicate parton absorption, or the time-reversed process. For example, PDFs whose momentum fractions lie in the sector with $x_2>0$, $x_3>0$ and $x_1<0$ represent the interference between a quark-gluon state and a quark state, with each point giving a different momentum partition in the quark-gluon system. Figure \ref{fig:hex} shows this hexagonal domain and illustrates the partonic interpretation of each sector. 

In addition to the two quark PDFs $T$ and $\Delta T$ (for each quark flavor), there are two independent gluon PDFs $T_{3F}^\pm$, defined by three-gluon operators, whose explicit definitions are given in ref.~\cite{Braun:2009mi, Rodini:2024usc}. This set of PDFs forms a closed system, in the sense that they mix with each other through the evolution equations but do not mix with other PDFs \cite{Bukhvostov:1985rn, Kodaira:1997ig, Braun:2001qx}, and that any chiral-even operator of geometrical twist-three can be expressed in their terms using mathematical manipulations and QCD equation of motions \cite{Balitsky:1987bk, Geyer:1999uq}. Therefore, this set of functions forms a fundamental and irreducible block of twist-three dynamics, similarly to unpolarized quark and gluon PDFs of twist-two. 

The evolution equations for PDFs of twist-three have the form
\begin{eqnarray}\label{ev-eq}
\frac{\partial \vec T(x_1,x_2,x_3;\mu)}{\partial \ln \mu}=[\mathbf{H}\otimes \vec T](x_1,x_2,x_3;\mu),
\end{eqnarray}
where $\vec T=(T,\Delta T,T_{3F}^+,T_{3F}^-)$, $\mathbf{H}$ is the matrix of integral kernels, and $\otimes$ is an integral convolution. The integral convolution is such that evolution of $\vec T$ with $(x_1,x_2,x_3)$ located at the dashed lines (see fig.~\ref{fig:hex}), involves the area between the dashed line and the boundary of the hexagon. Currently, the evolution kernel $\mathbf{H}$ is known at LO \cite{Bukhvostov:1985rn, Braun:2009mi}. The numerical solution of (\ref{ev-eq}) for a given boundary value is realized in ref.~\cite{Rodini:2024usc}, where further details on definitions and properties of twist-three PDFs can be found.

As this is the first determination of twist-three PDFs, there are no prior constraints on the functional form of the input ansatz. Therefore, we adopt a minimal parameterization that respects all required symmetries and vanishes at the boundary. Finer structural variations are captured through a polynomial dependence. In total, our ansatz contains 17 parameters: 3 common to all functions, 2 specific for the gluon parts, and the rest equally split between the $u$, $d$ and $s$ flavors. Notice, that twist-three PDFs are not separated into quark and anti-quark components, but present a single function for all combinations. Since most observables exhibit only weak sensitivity to $\Delta T$ and gluon distributions, these components are parametrized with fewer free parameters. Nevertheless, the QCD evolution induces mixing among the various components, leading to non-trivial shapes even from a relatively simple initial ansatz. The explicit expression for the parametrization is given in appendix B.

The ansatz is imposed at the initial scale $Q=1$GeV. This scale is chosen to lie below experimental data points so that several iterations of the evolution procedure can be applied before comparing with measurements, thus ensuring correlated parameter constraints. The starting distributions for the $c$ and $b$ quarks are set to zero at the initial scale, but they are generated dynamically through the QCD evolution at scales above their mass thresholds. To cross-validate our model we performed several fits using other types of anzatz, changing polynomial parametrization to trigonometric and logarithmical structures. These tests demonstrated very similar behavior to the model that was finally adopted.

{\it Processes and the data --- } To reconstruct the multi-dimensional shape of twist-three PDFs, one should consider different kinds of observables, such that each of them provides complementary constraining.  In our analyses we have studied the following observables:

\textbf{The structure function $g_2(x,Q^2)$}, which was measured in the polarized DIS at SLAC, HERMES and JLab \cite{E154:1997eyc, E155:1999eug, E155:2002iec, HERMES:2011xgd, JeffersonLabHallA:2004tea, JeffersonLabHallA:2016neg}. For our analyses we selected data points with $Q^2>2$GeV$^2$, and discarded other measurements such as \cite{E142:1996thl, E143:1995xmc,E143:1998hbs, SpinMuonSMC:1997mkb, RSS:2006tbm} due to their excessive uncertainty\footnote{We have checked that our final predictions describe the data by \cite{E142:1996thl, E143:1995xmc,E143:1998hbs, SpinMuonSMC:1997mkb, RSS:2006tbm} far within their uncertainty bands.}. The expression for $g_2(x)$ is proportional to the integral the values of $T$ and $\Delta T$ (\ref{def:g2-tw3}) over the regions shown by red and green in fig.~\ref{fig:hex}.

\textbf{The moment $d_2(Q^2)$}, which was measured in the polarized DIS (alongside with function $g_2$) or directly by QCD lattice simulations \cite{Burger:2021knd, Crawford:2024wzx}. The moment $d_2$ represents the integral of the twist-three PDF $T$ over the whole hexagon (\ref{def:d2}). The moment $d_2$ is a very small quantity, and the majority of measurements do no distinguish it from zero due to a very large uncertainty. Thus we selected measurements by \cite{JeffersonLabHallA:2016neg, HERMES:2011xgd, SANE:2018pwx, Burger:2021knd} with $Q^2>2$GeV$^2$. It is noteworthy mentioning that lattice measurements \cite{Burger:2021knd} provide values directly for $u$ and $d$ quarks, thus facilitating the flavor separation.

\textbf{The single-spin asymmetry $A_{UT}^{\sin(\phi_h-\phi_s)}$} in SIDIS, where the twist-three PDF enters via the Sivers function $f_{1T}^\perp$ \cite{Sivers:1989cc}. The description of SIDIS is made within the transverse momentum dependent (TMD) factorization approach \cite{Collins:2011zzd, Angeles-Martinez:2015sea, Echevarria:2014xaa}. The expression for the Sivers function is dominated \cite{Scimemi:2019gge, Ji:2006ub, Kanazawa:2015ajw, Scimemi:2018mmi} by the Qiu-Sterman matrix element $T(-x,0,x)$  (\ref{siv=T}) \cite{Efremov:1984ip, Qiu:1991pp, Bury:2021sue} and the range of its definition is shown by the red line in fig.~\ref{fig:hex}. The data for $A_{UT}^{\sin(\phi_h-\phi_s)}$ is provided by HERMES, COMPASS and JLab \cite{COMPASS:2008isr, COMPASS:2016led, HERMES:2020ifk, JeffersonLabHallA:2014yxb}. In addition to the cut $Q^2>2$GeV$^2$, we applied the cut $p_\perp/(zQ)<0.3$, to justify the application of TMD factorization.

\textbf{The double-spin asymmetry $A_{LT}^{\cos(\phi_h-\phi_s)}$} in SIDIS, where the twist-three PDF enters via the worm-gear-T TMD distribution $g_{1T}^\perp$, also known as the Kotzinian-Mulders function \cite{Kotzinian:1995cz}. Analogously to $A_{UT}^{\sin(\phi_h-\phi_s)}$, the description is based on the TMD factorization approach. The expression for the $g_{1T}^\perp$ is dominated by the integral convolution of $T$ and $\Delta T$ \cite{Rein:2022odl, Scimemi:2018mmi}. Structurally this convolution resembles that of $g_2$ (i.e. involves the red and green areas in fig.~\ref{fig:hex}) but has a different integral kernel. The data for $A_{LT}^{\cos(\phi_h-\phi_s)}$ is provided by HERMES and COMPASS \cite{COMPASS:2008isr, COMPASS:2016led, HERMES:2020ifk}, and are subjected to the same cuts as for $A_{UT}^{\sin(\phi_h-\phi_s)}$.

The explicit expressions for these observables are presented in appendix A. The computation of $g_2$, $A_{UT}^{\sin(\phi_h-\phi_s)}$ and $A_{LT}^{\cos(\phi_h-\phi_s)}$ involves also PDFs of twist-two, which we computed using state-of-the-art ingredients. These are the Wandzura-Wilczek contribution \cite{Wandzura:1977qf} to $g_2$ and to $g_{1T}^\perp$, which are computed using the helicity PDF set MAPPDFpol1.0 \cite{Bertone:2024taw}, and the unpolarized TMDPDF and TMD fragmentation function, which are taken from the ART25 extraction \cite{Moos:2025sal}. Note, that the TMD factorization part is evaluated using the ART25 setup, which is characterized as N$^4$LL. 

In total, our data set consists of 243 data points (see summary in table \ref{tab:chi}). Data selection followed strict criteria of theoretical applicability and our previous experience of description of spin-asymmetries \cite{Horstmann:2022xkk, Bury:2020vhj, Bury:2021sue}. The flavor separation is achieved mainly through the available data for SIDIS with $\pi^\pm$ and $K^\pm$ in the final state.

The present data set is sufficient for a proof-of-concept determination. Although it represents the largest compilation of twist-three data ever considered, this data set is not exhaustive. Many other processes are sensitive to twist-three physics and could be incorporated into the global analysis. Notable candidates are the transverse single-spin asymmetry $A_N$ in proton–proton collisions \cite{E581:1991eys, STAR:2003lxu} and twist-three effects in deeply virtual Compton scattering \cite{Belitsky:2000vx}.

\begin{table}[t]
\centering
\begin{tabular}{||l|c|c|c||}
Data set& References & $N_{\text{pt}}$ & $\chi^2/N_{\text{pt}}$  
\\\hline
\Xhline{5\arrayrulewidth}
$d_2$ & 
\cite{JeffersonLabHallA:2016neg, HERMES:2011xgd, SANE:2018pwx, Burger:2021knd}   & 7 & 0.99
\\
\Xhline{5\arrayrulewidth}
$g_2$  & & 103 & 0.98
\\
\Xhline{5\arrayrulewidth}
\multicolumn{1}{!{\vrule width 6pt}l|}{ E154} &  \cite{E154:1997eyc}   & 15 & 1.05
\\\hline 
\multicolumn{1}{!{\vrule width 6pt}l|}{ E155} &  \cite{E155:1999eug, E155:2002iec}   & 46 & 1.05
\\\hline 
\multicolumn{1}{!{\vrule width 6pt}l|}{ HERMES} &  \cite{HERMES:2011xgd}   & 13 & 0.98
\\\hline 
\multicolumn{1}{!{\vrule width 6pt}l|}{ Hall A} &  \cite{JeffersonLabHallA:2004tea, JeffersonLabHallA:2016neg}   & 29 & 0.82
\\
\Xhline{5\arrayrulewidth}
$A_{UT}$& & 63 & 1.17
\\
\Xhline{5\arrayrulewidth}
\multicolumn{1}{!{\vrule width 6pt}l|}{ COMPASS} &  \cite{COMPASS:2008isr, COMPASS:2016led}   & 16 & 0.51
\\\hline 
\multicolumn{1}{!{\vrule width 6pt}l|}{ HERMES} &  \cite{HERMES:2020ifk}   & 44 & 1.47
\\\hline 
\multicolumn{1}{!{\vrule width 6pt}l|}{ JLab} &  \cite{JeffersonLabHallA:2014yxb}   & 3 & 0.34
\\
\Xhline{5\arrayrulewidth}
$A_{LT}$& & 70 & 0.86
\\
\Xhline{5\arrayrulewidth}
\multicolumn{1}{!{\vrule width 6pt}l|}{ COMPASS} &  \cite{COMPASS:2008isr, COMPASS:2016led}   & 26 & 0.39
\\\hline 
\multicolumn{1}{!{\vrule width 6pt}l|}{ HERMES} &  \cite{HERMES:2020ifk}   & 44 & 1.14
\\
\Xhline{5\arrayrulewidth}
Total & & 243 & 0.99
\\\hline
\end{tabular}
\caption{\label{tab:chi} List of the values of $\chi^2$ obtained for set and subsets of data used in our analyses.}
\end{table}

{\it Results of extraction --- } The agreement between data and theory is quantified with a standard $\chi^2$-test function, constructed according to the standard procedure \cite{Ball:2008by} (see also \cite{Bertone:2019nxa, Horstmann:2022xkk, Bury:2021sue}). The analysis was performed using well-established methods, in part repeating procedures developed in our previous works \cite{Bury:2020vhj, Rein:2022odl}.

A central aspect of the analysis is the treatment and propagation of uncertainties. We employ the replica (parametric-bootstrap) method \cite{Ball:2008by}, in which random pseudo-data sets are generated according to the estimated probability distributions of all inputs to the $\chi^2$-function. In our work we distinguish and take into account two principal sources of uncertainty. \textit{Experimental uncertainties:} statistical and systematic uncertainties as reported by the experimental collaborations. As input to the parametric bootstrap, artificial data sets are generated following the procedure of ref.~\cite{Ball:2008by};
\textit{Theoretical uncertainties:} the uncertainties in twist-two PDFs that enter the description of observables. As input to the parametric bootstrap, we randomly select input replicas for both the helicity PDFs \cite{Bertone:2024taw} and the unpolarized TMDPDFs \cite{Moos:2025sal} for each minimization.

In total, we generated 300 pseudo–data/theory combinations and performed a least-squares minimization for each, producing a distribution of best-fit parameters used to compute secondary observables and plots. The average value of extracted twist-three PDFs provides $\chi^2/N_{\text{pt}}=0.99$, which points to an excellent description of the data. The table \ref{tab:chi} lists the values of $\chi^2$ obtained for individual data sets.

\begin{figure*}
\begin{center}
\includegraphics[width=0.65\textwidth]{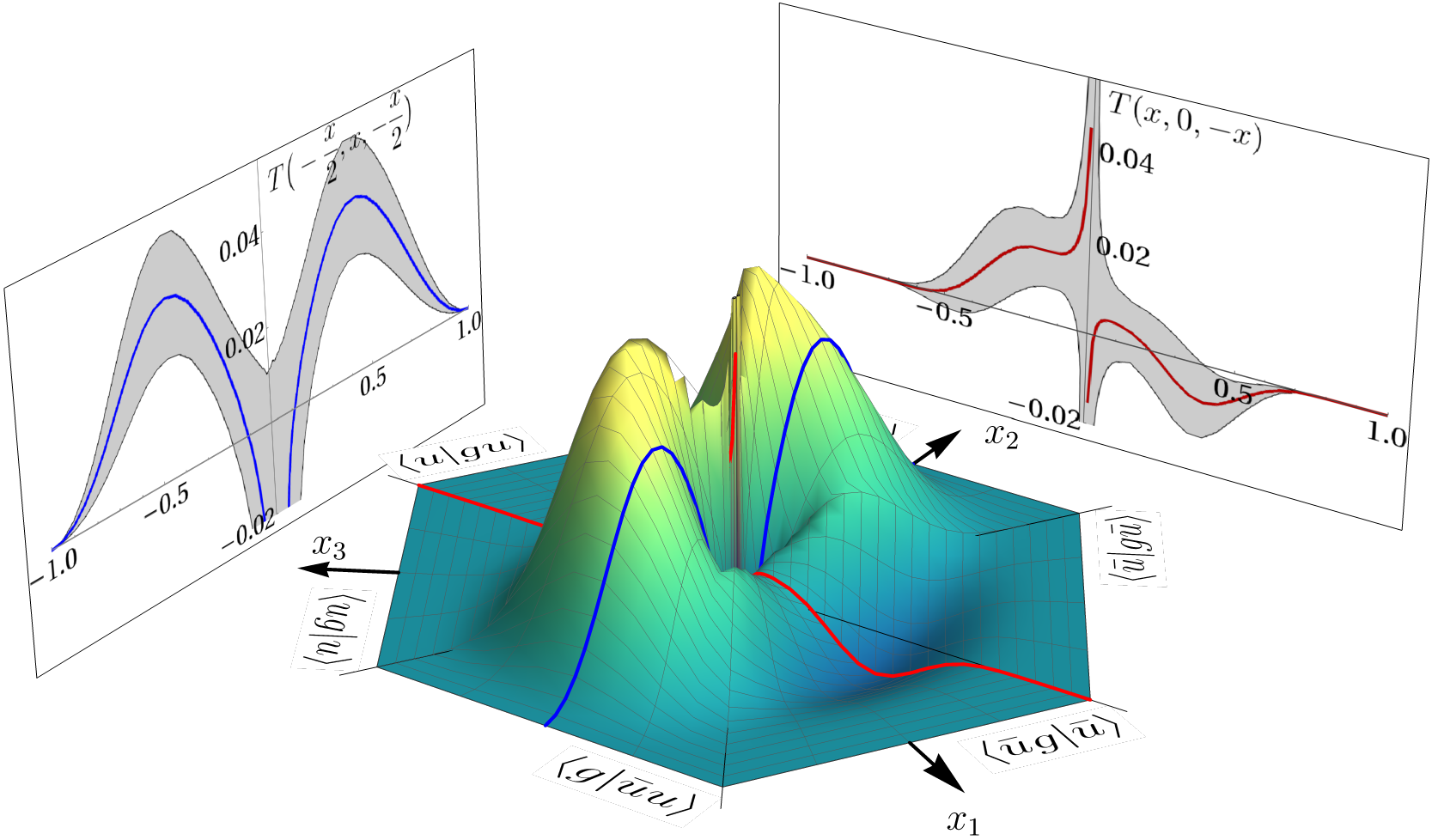}
\caption{\label{fig:u-quark} Twist-three PDF for $u$ quark  at 4GeV determined in this work. The uncertanties of the extraction (68\%CI) can be understood from the panels showing the sections of function by planes $x_2=0$ (red line) and $x_3=x_1$ (blue line).}
\end{center}
\end{figure*}

The evidence for a non-zero signal can be assessed against the null hypothesis that all twist-three PDFs vanish. Under this assumption $\chi^2/N_{\text{pt}}=1.38$, indicating a general similarity to the data. However, twist-three distributions are not positive definite, so the null hypothesis is, in fact, valid for some part of the data. Additionally, several experimental subsets carry very large uncertainties. Hence, it is more conclusive to examine only those data sets that themselves disfavor the null hypothesis, in particular we restrict to data sets with $\chi^2/N_{\text{pt}}>1$. For this reduced data set we found $\chi^2_{\text{null}}/N_{\text{pt}}=1.72$ (with $N_{pt}=152$), which improves to $\chi^2/N_{\text{pt}}=1.23$ after the fit. In other words, the combined analysis produces a marked and statistically meaningful improvement in the global description of the data. We emphasize that none of the considered observables can constrain our distributions on their own. This is not only due to a limited sensitivity of individual observables, but also because of inconclusive data quality. For example, the datasets for $g_2$ and $g_{1T}^\perp$ do not exclude the null hypothesis even at the $1\sigma$ level.

We also observed that, after the optimization of parameters, the distribution of $\chi^2$ values between data points became more uniform with generally $\chi^2/N_{\text{pt}}<1.2$ per data set, with  the sole exception of $K^+$-production measurements made at HERMES ($\chi^2/N_{\text{pt}}=2.33$ for $N_{\text{pt}}=11$). The replica ensembles for both parameters and $\chi^2$ show no anomalous features. Taken together, these tests demonstrate that the extracted signal and its quoted uncertainties provide a robust and reliable representation of the underlying physics.

Our analysis shows a decisive breakthrough: for the first time, we observe a clear and statistically significant signal of twist-three effects. The uncertainty bands are sufficiently narrow, so much so that in several regions, the values differ from zero at the $2\sigma-3\sigma$ level. Figure~\ref{fig:u-quark} illustrates the shape and its precision for the distribution $T$ of the $u$ quark, with similar accuracy found across other distributions. The values of fitted parameters, as well as illustrations for shapes of other distributions, are presented in the appendix B.

The magnitude of the distributions is on the order of $(2-5) \times 10^{-2}$. It is in general agreement with model estimations \cite{Lu:2004au, Braun:2011aw}. These values are roughly two orders of magnitude smaller than those of unpolarized PDFs, about one order of magnitude smaller than helicity PDFs for valence quarks, but comparable to helicity PDFs for sea quarks \cite{Ethier:2020way}. Generally, all flavors share a similar magnitude (except for heavy flavors that are generated perturbatively) consistently with the expectation that interference effects do not distinguish between valence and non-valence quarks.

The shapes of the distributions appeared to be rather non-trivial (see fig.\ref{fig:quarks} in appendix B). It is interesting to observe that the twist-three functions exhibit clearly distinct behavior across different sectors, despite being smooth and despite the input parameterization treating all sectors equally. We find that the $ug$ and $dg$ interference terms have opposite signs, consistent with predictions from various models \cite{Yuan:2003wk, Courtoy:2008vi}. Beyond this, our results cannot be directly compared with previous attempts to determine twist-three elements, since it is the first determination of its kind. All the observed emerging structures await deeper understanding and interpretation in future studies.

It is worth emphasizing that each observable considered in this work has a long history of consideration and represents an interesting part of QCD. Their joint consideration is a new word in the global QCD analyses, which will help to restrict and synchronize various studies of the proton's internal structure. As a by-product, we received rich information of various aspects of parton dynamics, such as the most accurate determination of the Sivers function, which is a key ingredient for the proton's tomography in momentum space \cite{Boer:2011fh, Angeles-Martinez:2015sea, Bacchetta:2020gko, Bury:2020vhj}; the average transverse-momentum displacement of quarks \cite{Burkardt:2003yg} and the averaged transverse force acting on the quark \cite{Burkardt:2008ps}, which for $u$ and $d$ flavors appear to be (here at $\mu=4$GeV, and uncertainties represent 68\%CI)
\begin{align}\nn
&\langle \vec k_u\rangle= 9.5^{+6.9}_{-7.1}\text{MeV}, && \langle \vec k_d\rangle= -18.7^{+18.1}_{-17.5}\text{MeV};
\\\nn
&\langle \vec f\rangle_u=-22.8^{+8.2}_{-8.1}\text{MeV/fm}, && \langle \vec f\rangle_d=54.7^{+17.9}_{-17.9}\text{MeV/fm}.
\end{align}
These details, together with other side results, will be presented in a subsequent publication.

{\it Conclusion --- }  We have, for the first time, determined the twist-three PDFs, which represent quark-gluon-quark interference within the proton and underlie many perplexing effects in particle scattering. The extracted distributions exhibit well-controlled uncertainties and deviate from zero at the $2\sigma-3\sigma$ level, marking a clear and statistically significant observation. This level of precision was made possible only by the simultaneous analysis of four different types of observables, selected based on our previous experience, and due to incorporation of the complete set of evolution equations for twist-three PDFs. Looking ahead, incorporating additional observables into such analyses has the potential to improve accuracy further. 

The observation of internal parton-interference effects invites for a reassessment of the modern picture of the proton. For decades, a high-energy proton has been described solely in terms of probability densities. This view has become so ingrained that, in much of modern high-energy physics, a fast proton is effectively treated as a beam of classical particles. In this picture, the quantum nature of partons, beyond perturbative effects, is entirely disregarded. Our results show that interference contributions can be comparable in size to density contributions for polarized quarks. This implies that, internally, the proton is far more quantum than previously assumed, at least in its spin-dependent component. While the broader implications of this discovery are still unfolding, the identification of twist-three distributions opens an exciting new frontier in our understanding of the proton's internal dynamics.

\vspace*{0.3cm}
%%%%%%%%%%%%%%%%%%%%%%%%%%%%%%%%%%%%%%%%%%%%%%%%%%%%%%%%%%%%%%%%%%%%%%%%%%
{\it Acknowledgments ---}  We thank Vladimir Braun and Alexei Prokudin for useful comments and stimulating discussions. A.V. is funded by the Atracci\'on de Talento Investigador program of the Comunidad de Madrid (Spain) No. 2020-T1/TIC-20204. The work of S.R. is supported by the German Science Foundation (DFG), grant number 409651613 (Research Unit FOR 2926),  subproject 430915355. The project is supported by grants ``Europa Excelencia'' No. EUR2023-143460 funded by MCIN/AEI/10.13039/501100011033/ by the Spanish Ministerio de Ciencias y Innovaci\'on.  This project is also supported by the European Union Horizon research Marie Skłodowska-Curie Actions – Staff Exchanges, HORIZON-MSCA-2023-SE-01-101182937-HeI, DOI: 10.3030/101182937.

{\it Data availability ---} All experimental data analysed in this study are publicly available from the references cited in the text and in Table \ref{tab:chi}. The resulting twist-three PDF grids, and replica ensembles are deposited at https://github.com/VladimirovAlexey/artemide-public under DOI:  https://doi.org/10.5281/zenodo.17153216.

\clearpage
\begin{figure*}[t]
\begin{center}
\includegraphics[width=0.99\textwidth]{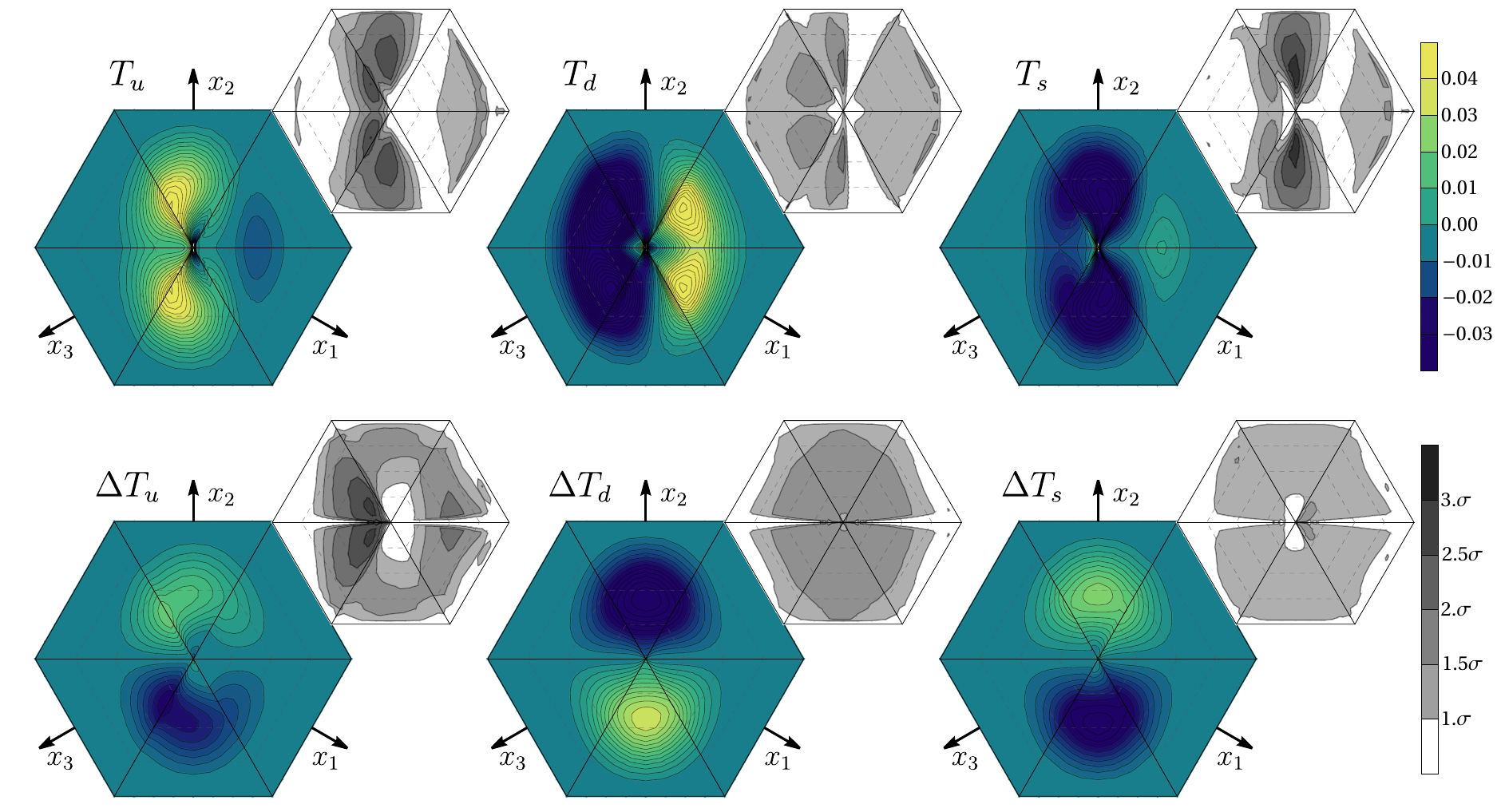}
\caption{\label{fig:quarks} The mean value of twist-three PDFs for $u$, $d$ and $s$ quarks at 4GeV. The uncertainty bands are demonstrated in smaller gray plots as the ratio of the size of the uncertainty band to the mean value measured in $\sigma$'s. I.e. the white regions imply that the zero value is covered by the uncertainty band, while darker regions do not contain the zero value within uncertainty bands.}
\end{center}
\end{figure*}

\begin{widetext}
\section{Appendix A: Expressions for observables}
\label{app}

In this appendix we present the expressions for cross-sections, and structure functions used in our analyses. 

The structure function $g_2$ at LO is given by \cite{Jaffe:1989xx, Ji:1990br, Kodaira:1994ge, Kodaira:1996md, Ali:1991em}
%\begin{widetext}
\begin{eqnarray}\label{def:g2}
&&g_2(x,Q^2)=-g_1(x,Q^2)+\int_x^1 \frac{dy}{y}g_1(y,Q^2)
+\sum_f \frac{e_f^2}{2} \int_x^1 \frac{dy}{y}\(\Delta q_T(y,Q^2)+\Delta q_T(-y,Q^2)\),
\end{eqnarray}
where $Q$ is the DIS photon virtuality, $x$ is the Bjorken variable, $e_f$ is the electric charge of a quark with flavor $f$, and
\begin{eqnarray}\label{def:g2-tw3}
\Delta q_T(x,Q^2)=\int [d\xi]\frac{T(\xi_1,\xi_2,\xi_3,Q^2)+\Delta T(\xi_1,\xi_2,\xi_3,Q^2)}{M} \frac{d}{d\xi_3}\frac{\delta(x+\xi_1)-\delta(x-\xi_3)}{\xi_1+\xi_3}.
\end{eqnarray}
%\end{widetext}
The first two terms of $g_{2}$ in (\ref{def:g2}) represent the Wandzura-Wilczek contribution \cite{Wandzura:1977qf}. These are twist-two terms, which were computed using the helicity PDF set MAPPDFpol1.0 \cite{Bertone:2024taw}. The integral convolution in (\ref{def:g2-tw3}) is such that for a given $x$ it integrates over the regions shaded by red (for $\Delta q_T(y)$) and by green (for $\Delta q_T(-y)$) in fig.~\ref{fig:hex}. 

The moment $d_2$ is defined as \cite{Jaffe:1989xx}
\begin{eqnarray}\label{def:d2}
d_2(Q^2)&=&3\int_0^1 dx x^2 \bar g_2(x,Q^2)
%\\\nn &=&
=
\sum_{f}\frac{e_f^2}{2}\int [d\xi]T(\xi_1,\xi_2,\xi_3,Q^2),
\end{eqnarray}
where $\bar g_2$ is the pure twist-three part of $g_2$ (given by the last term of (\ref{def:g2})). Therefore, the moment $d_2$ represents the integral over the whole hexagonal domain, and provides the normalization of the twist-three PDF $T$.

The spin-asymmetries in SIDIS are defined as the ratio of certain structure functions \cite{Bacchetta:2004jz, Bacchetta:2006tn} 
\begin{eqnarray}\nn
A_{UT}^{\sin(\phi_h-\phi_S)}(x,z,p_\perp,Q)&=&\frac{F_{UT}^{\sin(\phi_h-\phi_S)}}{F_{UU,T}},
%\\
\qquad
A_{LT}^{\cos(\phi_h-\phi_S)}(x,z,p_\perp,Q)=\frac{F_{LT}^{\cos(\phi_h-\phi_S)}}{F_{UU,T}},
\end{eqnarray}
where $x$, $z$, $p_\perp$, and $Q$ are the standard kinematic variables of SIDIS. In turn, the structure functions are expressed via the convolution of TMD distributions as \cite{Boer:2011xd}
%\begin{widetext}
\begin{eqnarray}
F_{UU,T}&=&\Big|C_V\(\frac{Q}{\mu}\)\Big|^2\sum_f e_f^2 \int_0^\infty \frac{b\,db}{2\pi}J_0\(\frac{b|p_\perp|}{z}\)f_{1,f}(x,b;\mu,Q)d_{1,f}(z,b;\mu,Q),
\\
F_{UT}^{\sin(\phi_h-\phi_S)}&=&\Big|C_V\(\frac{Q}{\mu}\)\Big|^2\sum_f e_f^2 \int_0^\infty \frac{b^2\,db}{2\pi}J_1\(\frac{b|p_\perp|}{z}\)f_{1T,f}^\perp(x,b;\mu,Q)d_{1,f}(z,b;\mu,Q),
\\
F_{LT}^{\cos(\phi_h-\phi_S)}&=&\Big|C_V\(\frac{Q}{\mu}\)\Big|^2\sum_f e_f^2 \int_0^\infty \frac{b^2\,db}{2\pi}J_1\(\frac{b|p_\perp|}{z}\)g_{1T,f}^\perp(x,b;\mu,Q)d_{1,f}(z,b;\mu,Q),
\end{eqnarray}
where $C_V$ is the hard coefficient function, $J_0$ and $J_1$ are the Bessel functions, and $f_1$, $d_1$, $f_{1T}^\perp$ and $g_{1T}^\perp$ are the unpolarized TMDPDF, TMDFF, Sivers function, and worm-gear-T function, respectively. The perturbative elements of these expression are taken according to the ART25 setup \cite{Moos:2025sal}, which is summarized as N$^4$LL. The definition of evolution scales is made within the $\zeta$-prescription \cite{Scimemi:2018xaf, Scimemi:2019cmh}. The values of the unpolarized parts are defined in the ART25 fit \cite{Moos:2025sal}.

The relation of Sivers and worm-gear-T functions to twist-three PDF is established in \cite{Scimemi:2019gge, Ji:2006ub, Kanazawa:2015ajw, Scimemi:2018mmi, Rein:2022odl}. At LO it reads
\begin{eqnarray}\label{siv=T}
&&\lim_{b\to 0}f_{1T}^\perp(x,b)=-\pi T(-x,0,x;\mu_{\text{OPE}})+\mathcal{O}(\alpha_s),
\\
\label{wgt=T}
&&\lim_{b\to 0}g_{1T}^\perp(x,b)=x\int_x^1\frac{dy}{y}g_1(y,\mu_{\text{OPE}})
+
2x\int [d\xi]\int_0^1 d\alpha \delta(x-\alpha \xi_3)\(\frac{\Delta T}{\xi_2^2}+\frac{T-\Delta T}{2\xi_2\xi_3}\)
,
\end{eqnarray}
\end{widetext}
where the argument of $T$ and $\Delta T$ is $(\xi_1,\xi_2,\xi_3,\mu_{\text{OPE}})$. Here $\mu_{\text{OPE}}=\frac{2e^{-\gamma_E}}{|b|}+5\text{GeV}$ is the scale of the operator product expansion adapted from ART25.
%\begin{eqnarray}\label{muOPE}
%\mu_{\text{OPE}}=\frac{2e^{-\gamma_E}}{|b|}+5\text{GeV}.
%\end{eqnarray}
The sign in (\ref{siv=T}) corresponds to the SIDIS definition of the Sivers function. This information is incorporated into the ansatz for TMD distribution as %(for Sivers function)
\begin{eqnarray}
f_{1T}^\perp(x,b)=-\pi T(-x,0,x;\mu_{\text{OPE}})f_{\text{NP}}(x,b),
\end{eqnarray}
where $f_{\text{NP}}$ is a nonperturbative shape function such that $f_{\text{NP}}\sim 1+b^2...$ at small-$b$. The function $g_{1T}^\perp$ has an analogous expression. The expression for $f_{\text{NP}}$ is taken similar to those of unpolarized distributions \cite{Moos:2025sal, Moos:2023yfa} $f_{\text{NP}}=\cosh^{-1}(\lambda b)$, with $\lambda$ being a free parameter.
%\end{widetext}

\section{Appendix B: Extracted distributions}

We adopt a minimal parameterization of twist-three PDFs that respects all required symmetries \cite{Braun:2009mi, Rodini:2024usc} and vanishes at the boundaries of the kinematic domain. The $f=u, d, s$ flavours are treated independently, each with the same ansatz
\begin{align}
T_f(x_1,x_2,x_3)&=h(x_1,x_2,x_3)\\\nn &\times  \left[ \alpha^f_1+\alpha_2^f(x_1-x_3) + \alpha^f_2 x_1 x_3 \right],
\\
\Delta T_f(x_1,x_2,x_3)&=h(x_1,x_2,x_3) \alpha^f_4 x_2,
\end{align}
where $\alpha^f_i$ are free parameters. The common envelope function is
\begin{equation}
h(x_1,x_2,x_3) = \frac{(1-x_1^2)^a (1-x_2^2)^b (1-x_3^2)^a}{(x_1^2 + x_2^2 + x_3^2)^c},
\end{equation}
and has $\{a,b,c\}$ as free parameters, constrained by $a,b>1$ and $c<1$. Gluon distributions is presented as:
\begin{align}
T_{3F}^+(x_1,x_2,x_3) &= \beta_1 (x_1-x_3)h(x_1,x_2,x_3),\\
T_{3F}^-(x_1,x_2,x_3) &= \beta_2 h(x_1,x_2,x_3),
\end{align}
where $\beta_{1,2}$ are free parameters. Therefore, the model contains a total of 17 free parameters. The function $h$ controls the center and boundary  behavior, while the parameters $\alpha$ and $\beta$ shape the interior part.

The fitting procedure yields the following values
\begin{eqnarray}\nn
a&=&6.0_{-0.4}^{+0.3},\quad b=1.03_{-0.03}^{+0.03},\quad c=-1.48_{-0.05}^{+0.09},
\end{eqnarray}
%while those for the ansatz read:
\begin{align}\nn
\alpha_1^u&=1.2_{-0.3}^{+0.2},
&\alpha_2^u&=0.58_{-0.62}^{+0.57},
\\\nn 
\alpha_3^u&=8.3_{-2.4}^{+0.6},
&\alpha_4^u&=3.0_{-0.9}^{+0.5},
\\
\alpha_1^d&=-0.54_{-0.07}^{+0.08},
&\alpha_2^d&=1.3_{-1.1}^{+0.5},
\\ \nn
\alpha_3^d&=-10._{-2.}^{+4.},
&\alpha_4^d&=-22._{-3.}^{+6.},
\\\nn
\alpha_1^s&=-1.3_{-0.1}^{+0.3},
&\alpha_2^s&=-8.9_{-0.9}^{+3.1},
\\\nn 
\alpha_3^s&=4.1_{-1.7}^{+0.4},
&\alpha_4^s&=1.2_{-1.1}^{+0.4},
\\\nn
\beta_1&=-2.7_{-1.0}^{+1.4},
&\beta_2&=2.1_{-1.7}^{+0.8}.
\end{align}
The uncertainty represent as 68\%CI determined by bootstrap method. The shape of extracted distributions, as well as, the deviation of PDF from zero counted in the number of $\sigma$'s, is shown in fig.\ref{fig:quarks}.

\bibliography{bibliography}

\end{document}